\documentclass[aps,prl,preprint,groupedaddress]{revtex4-1}

\usepackage{graphicx}
\usepackage{dcolumn}
\usepackage{bm}
\begin{document}


\title{Positron acceleration by plasma wake fields driven by a hollow
electron beam}


\author{Neeraj Jain}
\thanks{Work done at Institute for Research in Electronics and Applied Physics,
 University of Maryland, College Park, MD, USA and supported by the US DoE.}
\affiliation{Max Planck Institute for Solar System Research,
Justus-von-Liebig-Weg 3, 37077, G\"ottingen, Germany}

\author{T. M. Antonsen Jr.}
\affiliation{Institute for Research in electronics and Applied Physics,
 University of Maryland, College Park, MD, USA}
\author{J. P. Palastro}
\thanks{Work done at Institute for Research in Electronics and Applied Physics,
 University of Maryland, College Park, MD, USA and supported by the US DoE.}
\affiliation{Icarus Research, Inc., P.O. Box 30780, Bethesda, Maryland
20824-0780}%


\date{\today}

\begin{abstract}
A scheme of wake field generation for positron 
acceleration using hollow or donut shaped electron driver beams is studied. An
annular
shaped,  electron free region forms around a hollow driver beam creating 
a 
favorable region (longitudinal field is accelerating and transverse field is focusing and radially linear)
for positron acceleration. Accelerating gradients of the order of 10 GV/m are
produced by a hollow electron beam driver with FACET like parameters. The peak
accelerating field increases linearly with the total charge in the beam driver
while the axial size of the favorable
region
($\sim$ one plasma wavelength) remains approximately fixed. The radial size
drops with the total charge but remains large enough for the placement of a
witness positron beam. We simulate an efficient acceleration of a 23 GeV
positron beam to 35.4 GeV with a maximum energy spread of 0.4\% and very
small emittance over a plasma length of 140 cm. 
\end{abstract}

\pacs{}

\maketitle

The plasma based
particle acceleration schemes, first proposed in 1979 \cite{tajima1979}, have
already achieved  acceleration gradients ($\sim$ tens of GeV/m) much larger than
those ($\sim$ tens of MeV/m) in conventional radio frequency accelerators.
In plasma based particle acceleration schemes,
a high
intensity laser or an ultra-relativistic charged particle beam propagates
through
a plasma generating electromagnetic fields in its 
wake, known as wake fields. 
 In the so called blowout regime of Plasma Wake Field Acceleration (PWFA)
\cite{chen1985,rosenzweig1991,lu2006}, all the plasma electrons are
expelled from the
path of a short (approximately one plasma wavelength
$k_p^{-1}=c/\omega_{p}$ long) and dense (beam density $>$ plasma density)
electron beam driver forming an electron free region
known as a 
bubble or ion channel. The expelled electrons then fall back to the beam
propagation axis behind the driver.  The blowout regime offers
a nearly radially uniform high acceleration gradient for  efficient acceleration
of electrons. 
The electron acceleration to  high energies (energy
doubling of 43 GeV electrons in an 85 cm  long plasma) has been
demonstrated in PWFA experiments \cite{blumenfeld2007}. Plasma-based acceleration of positrons, on the other hand, has been less explored and is essential for successful operation of an electron-positron collider.  

Current positron acceleration schemes generate wake fields in plasma 
either by an electron \cite{lotov2007} or by a positron beam driver
\cite{lee2001}.
 When the wake fields are driven by an electron beam driver in the bubble regime,  the
favorable region for positron acceleration (transverse field focusing and
longitudinal field accelerating)  forms between the first and second bubble. The favorable region has a narrow extent between the two
bubbles. The accelerating electric field varies rapidly with  axial
coordinate leading to large energy spread in the
accelerated positron beam (witness beam).
Furthermore, the transverse focusing field causes an increase in the emittance of the
witness beam.  
 In the case of a positron driver, the plasma electrons are attracted towards
rather
than blown out of the driver beam path and do not cross the axis in a narrow
region as they do in case of electron beam driver \cite{lee2001}. As a result
the accelerating fields are smaller than those driven by electron beams. 
The focusing fields are nonlinear in the transverse coordinate and vary along the axis of the beam
leading to the emittance growth of the witness beam. The
accelerating gradients can be improved if the positron beam driver propagates
through a hollow plasma channel \cite{lee2001}. For an appropriate hollow plasma channel radius, plasma electrons can cross the axis in a
narrow region increasing the wake field amplitude. Recently, self injection of
hollow electron bunch in the wake fields driven by a Laguerre-Gaussian laser
pulse and  positron acceleration was observed in simulations \cite{vleira2014}. 

In this Letter, we present a scheme of wake field generation for
positron 
acceleration using hollow or donut shaped electron drive beams (the beam density is
maximum at an off
axis location). The hollow electron beam pushes the plasma electrons towards its 
axis setting up the wake fields for positron acceleration in the hollow region.
 The accelerating field for positrons increases with
the total charge in the beam driver while the axial size of the favorable region
($\sim$ one plasma wavelength) remains approximately unchanged. This is in
contrast to the case of solid beam driver in which the size of the favorable
region diminishes with increasing charge in the beam. 

We calculate wake fields  driven by the propagation of
a hollow
electron beam in a uniform plasma in an azimuthally symmetric
($\partial/\partial\theta=0$) cylindrical geometry using the quasi-static code
WAKE
\cite{mora1997}. The quasi-static approximation exploits the disparity of 
driver and plasma evolution time scales. 
The time scale of evolution of the ultra-relativistic electron beam 
(relativistic factor $\gamma_b >> 1$) is the betatron period
$\tau_{b}=\sqrt{2\gamma_b}\lambda_p/c$ which is
much larger than the plasma time scale $\lambda_p/c$, where $\lambda_p$ is the
plasma wavelength. In the code WAKE, the response of the kinetic, warm and
relativistic plasma is calculated on a fast time scale assuming a fixed beam
driver. The
driver is then evolved over longer time scales \cite{jain2014}. 

We employ a moving computational domain which  changes its
axial position as the beam driver propagates along the axis. The axial coordinate ($\xi$) in the moving computational
domain can be written as $\xi=ct-z$. The initial number density of the hollow or donut
shaped electron beam driver is expressed as,
\begin{eqnarray}
 n_{beam}&=&n_b
\exp\left[-\frac{(r-r_0)^2}{2\sigma_r^2}-\frac{(\xi-\xi_0)^2}{2\sigma_z^2}
\right ]\label{eq:driver}
\end{eqnarray}
The peak number density ($n_b$) of the hollow beam is located at 
an off axis location $(r_0,\xi_0)$ and falls off within distances $\sigma_r$
(radially) and $\sigma_z$ (axially). The beam is completely hollow in the limit
$r_0/\sigma_r\rightarrow \infty$. Otherwise there is a small but finite density
in the core of the beam. In the limit $r_0\rightarrow 0$, the beam density has
peak at the axis and corresponds to a solid beam.
In our simulations we take $r_0\ge 4\sigma_r$ for hollow beams and $r_0=0$ for
solid beams. And thus,  we measure $r_0$ in terms of $\sigma_r$, i.e., we vary
$r_0/\sigma_r$ instead of $r_0$.

We first study wake field generation by the non-evolving electron driver. 
In the simulations, the plasma density is uniform $n_p=2\times 10^{17}$ cm$^{-3}$ (typical plasma density in FACET \cite{hogan2010})
giving $k_p^{-1}=c/\omega_{pe}=12 \, \mu$m. The beam is axially centered at
$\xi_0=0$.  The value of $\sigma_z=24\,\mu$m
$=2\, k_p^{-1}$ is fixed for all the results presented here for non-evolving
driver. We vary the
values of $\sigma_r$, $r_0/\sigma_r$ and the total charge $Q_d=-e\int
n_{beam}(r,\xi) r\,dr\,d\theta\,d\xi$ contained in the driver. 
The beam driver with an
initial energy 23 GeV is
modeled using $4\times 10^6$ simulation particles. The plasma is modeled using
9 particles per cell. The simulation domain size along $\xi$ is $15.6 \,
k_p^{-1}$ 
with a grid resolution of $0.01 \, k_p^{-1}$. The simulation domain size and
grid resolution in radial direction depends on the the values of $r_0$ and
$\sigma_r$. 

\begin{figure}
 \includegraphics[width=0.5\textwidth,height=0.4\textheight]
 {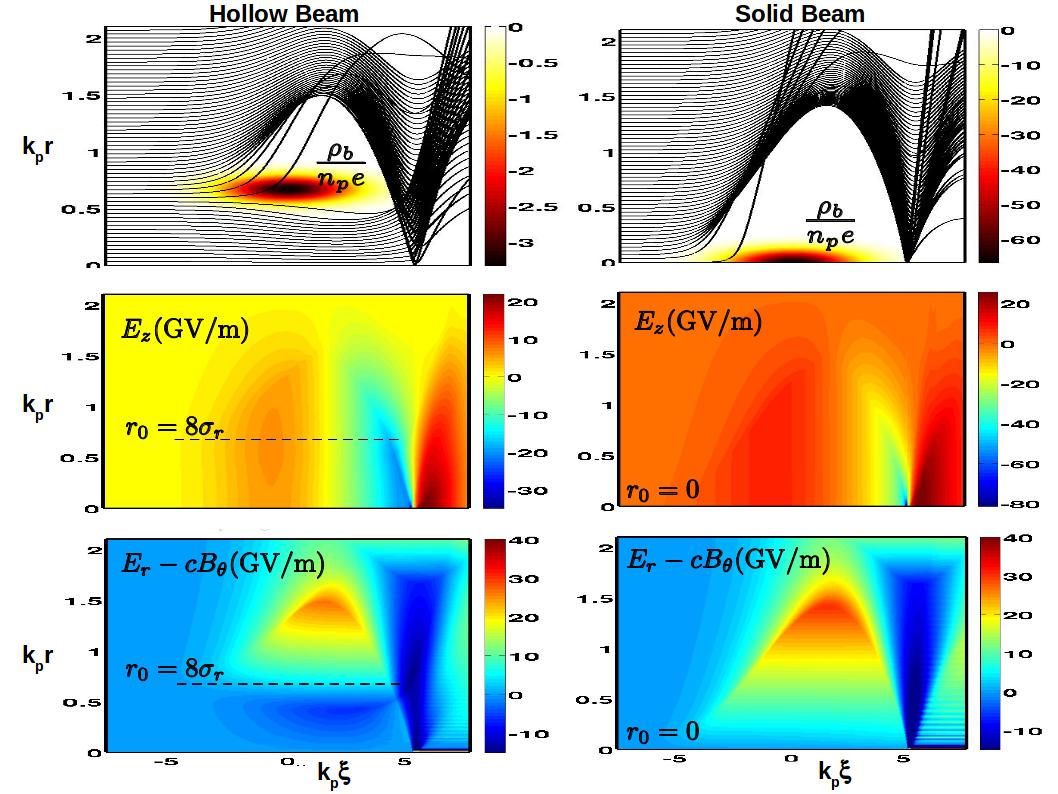}
 \caption{\label{fig:traj_solid_hollow}Trajectories of
plasma particles (black lines) and driver beam density (top), longitudinal electric field $E_z$ (middle) and transverse field $E_{rad}=E_r-cB_{\theta}$ (bottom) in the $r-\xi$ simulation domain for a hollow
($r_0=8\sigma_r$, left column)
and a solid ($r_0=0$, right column) electron beam drivers with
$\sigma_r=1\mu$m and $Q_d=0.8$ nC. 
 In case of the 
hollow beam, a horizontal dashed line marks
the radial location of the peak density of the beam driver. The color scale for
the transverse wake field is saturated at -15 GV/m in order to improve the
visibility of the negative (focusing for positrons) transverse field
below the dashed horizontal
line.}
\end{figure}

Plasma electrons follow different trajectories, shown in top panels of Fig.
\ref{fig:traj_solid_hollow}, in response to the hollow
($r_0=8\,\sigma_r$) and  solid ($r_0=0$) electron beam drivers both of which
have $\sigma_r=1\,\mu$m and $Q_d=0.8$ nC. In both cases, plasma
electrons are expelled from the beam. But since the bulk of the hollow beam is
centered
off axis, the plasma electrons are expelled both
towards and away from the axis. This is in contrast with the case of a solid
beam in which plasma electrons are expelled only away from the axis. The
plasma electrons move towards the axis through the hollow region  and
experience a force of repulsion due to other plasma electrons moving
towards the axis. For this reason, the radial deflection of
plasma trajectories in the hollow region is small until they intersect the back
of the bubble. At the back 
of the driver, plasma electrons are pulled back (towards their original radial
position)
 by the annular shaped electron free region.
 

The resulting structures of the wake fields are shown in the middle and bottom panels of Fig.
\ref{fig:traj_solid_hollow}. 
The longitudinal electric field $E_z$ is structurally the same for solid and
hollow
beam drivers. However, the structure of the transverse field
$E_{rad}=E_r-cB_{\theta}$
changes for hollow beams. As the value of $r_0$ becomes finite,  the
transverse
field in the hollow  region of the beam (below the horizontal dashed lines in
Fig. \ref{fig:traj_solid_hollow}) changes its direction from radially
outward (for $r_0=0$) to radially inward,
and thus, becomes focusing for positrons.    The reason for this change in the
case of a hollow beam is the formation of an annular shaped electron free region
which has positive charge density due to background ions.
Since the net charge density $e(n_p-n_e)$
below and above 
the electron free region is negative due to excess plasma electrons, 
the radial electric field points radially inward and outward near the bottom and
top boundaries of the electron free region, respectively. 

\begin{figure}
 \includegraphics[width=0.5\textwidth,height=0.3\textheight]
 {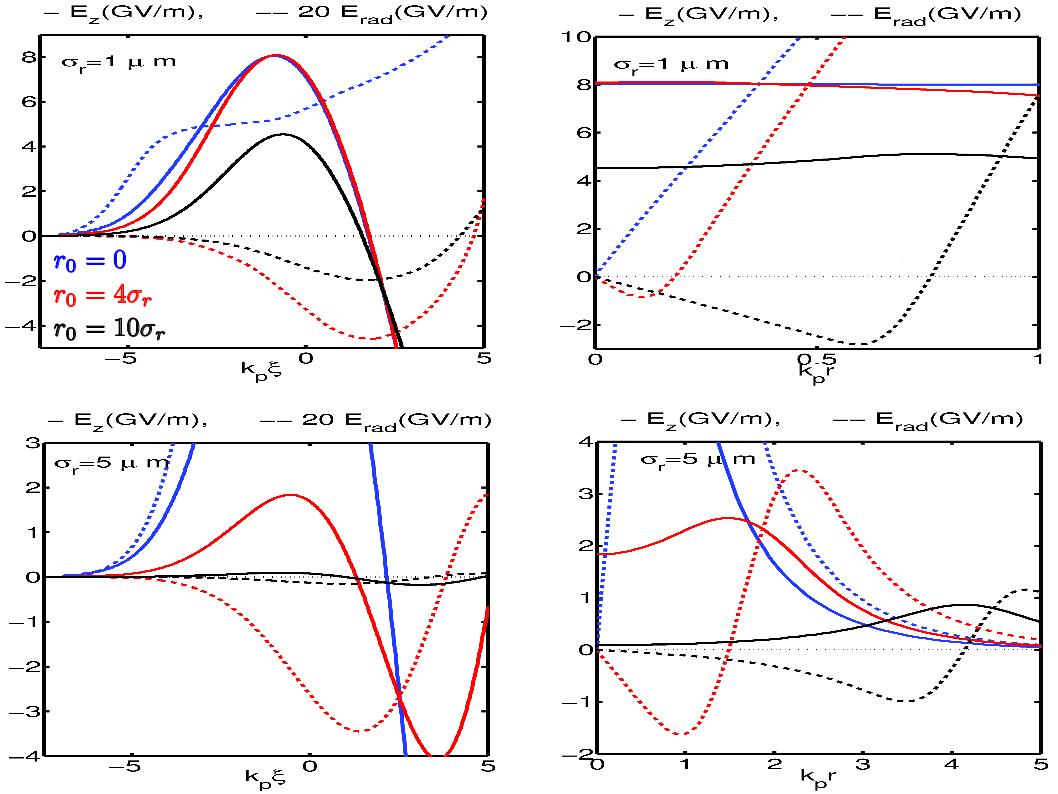}
 \caption{\label{fig:Ez_Erad_lineouts_sr1_1and5micron} Line-outs of Longitudinal
electric field $E_z$ and transverse field $E_{rad}=E_r-cB_{\theta}$ along the
axis
($r=0$; left column) and radius (at axial locations where $E_z$ is maximum in
the
favorable region for positron acceleration; right column) for
$\sigma_r=1\mu$m (top row) and $5\mu$m (bottom row), and for solid ($r_0=0$)
and hollow ($r_0=4\sigma_r$ and $10\sigma_r$)
electron beam drivers. The total charge in the beam driver is $Q_d=0.8$ nC} 
\end{figure}
 
\begin{figure}
 \includegraphics[width=0.5\textwidth,height=0.3\textheight]
 {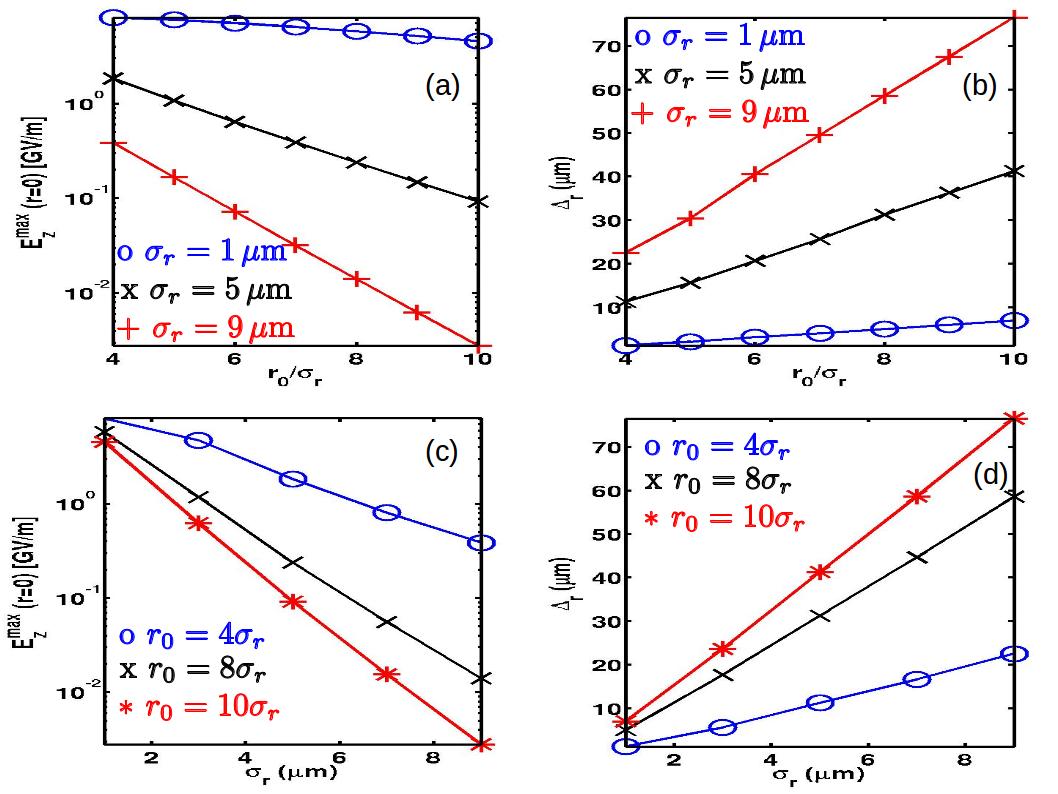}
 \caption{\label{fig:Ezmax_deltaR_r0_sr} Maximum 
on-axis longitudinal electric field $E_z^{max} (r=0)$ (left column)  and radial size $\Delta_r$ of the favorable region (right column) as  functions of $r_0/\sigma_r$ (top row) and $\sigma_r$ (bottom row) for various values of
$\sigma_r$ and $r_0/\sigma_r$. The total beam charge is $Q_d=0.8$ nC. The y-axes in the left column are logarithmic.} 
\end{figure}

The longitudinal electric field is positive in  part of the region where
the transverse force is focusing for positrons. The axial line-outs  of $E_z$
along the axis ($r=0$) and of 
$E_r-cB_{\theta}$ just above the axis, Fig.
\ref{fig:Ez_Erad_lineouts_sr1_1and5micron}, show that the axial size of the
favorable
region for positron acceleration (positive $E_z$ and negative $E_r-cB_{\theta}$)
is of the order of a plasma wavelength $\lambda_p=2\pi/k_p$ both for
$\sigma_r=1\,\mu$m and $\sigma_r=5\,\mu$m.  The axial size is slightly smaller
for $\sigma_r=5\,\mu$m. It depends very weakly
on the value of $r_0$.
For $\sigma_r=1\,\mu$m,  
$E_z$ in the favorable region is  
almost the same for the solid beam and the hollow beam with $r_0=4\sigma_r$ 
 however, it
drops for $r_0>4\sigma_r$. For
$\sigma_r=5\,\mu$m, $E_z$ has values much smaller than those for
$\sigma_r=1\,\mu$m and its drop with $r_0$ is also faster. The
logarithmic-variation of the maximum value of $E_z(r=0)$ in the favorable region
with $r_0/\sigma_r$ for a given $\sigma_r$, (Fig.
\ref{fig:Ezmax_deltaR_r0_sr}a) and with  $\sigma_r$ for a given $r_0/\sigma_r$ (Fig.
\ref{fig:Ezmax_deltaR_r0_sr}c) can be reasonably approximated as linear. This
indicates exponential dependence of $E_z^{max}(r=0)$ on $\sigma_r$ and
$r_0/\sigma_r$.

It is required for the emittance preservation of the witness beam
that the focusing field be linear in the radial direction and uniform in
the axial direction.
Although the transverse field in the favorable region (below the dashed
horizontal line in Fig. \ref{fig:traj_solid_hollow}) is negative (focusing for
positrons) until $r\approx r_0$, its radial variation can be approximated as
linear only before it reaches its negative peak as can be seen in radial
line-outs of $E_r-cB_{\theta}$ in Fig.
\ref{fig:Ez_Erad_lineouts_sr1_1and5micron}. For $\sigma_r=5\,\mu$m,
$E_r-cB_{\theta}$ becomes slightly nonlinear even before its negative peak
(curves are concave towards origin).
We define approximately the radial size $\Delta_r$ of the linearly varying
focusing
field as the radial distance at which  $d(E_r-cB_{\theta})/dr=0$. 
Figs. \ref{fig:Ezmax_deltaR_r0_sr}b and \ref{fig:Ezmax_deltaR_r0_sr}d show that
$\Delta_r$ increases linearly
with both of $r_0/\sigma_r$ (for fixed $\sigma_r$) and $\sigma_r$ (for fixed $r_0/\sigma_r$). The rate of increase with $\sigma_r$  depends on $r_0/\sigma_r$ and that with $r_0/\sigma_r$ on $\sigma_r$. In
the axial direction, the focusing field is not uniform but varies very slowly 
compared with that driven by a solid electron beam driver
or by a positron beam driver. 

Another desirable feature in plasma wake field acceleration is the axial
and radial uniformity of the accelerating wake field. Fig.
\ref{fig:traj_solid_hollow} and radial line-outs in Fig.
\ref{fig:Ez_Erad_lineouts_sr1_1and5micron} show that  $E_z$ is highly uniform
in the radial direction for $\sigma_r=1\,\mu$m. The radial uniformity of
$E_z$ is degraded for larger values of $\sigma_r=5\,\mu$m. Although the
magnitude of $E_z$ in the
favorable
region varies with $\xi$, the variation is not as rapid as near the
electric field spike resulting from a solid beam driver.  The radial uniformity
and
relatively slow variation with $\xi$ of $E_z$ are ideal for positron
acceleration with low energy spread.

\begin{figure}
 \includegraphics[width=0.5\textwidth,height=0.3\textheight]
 {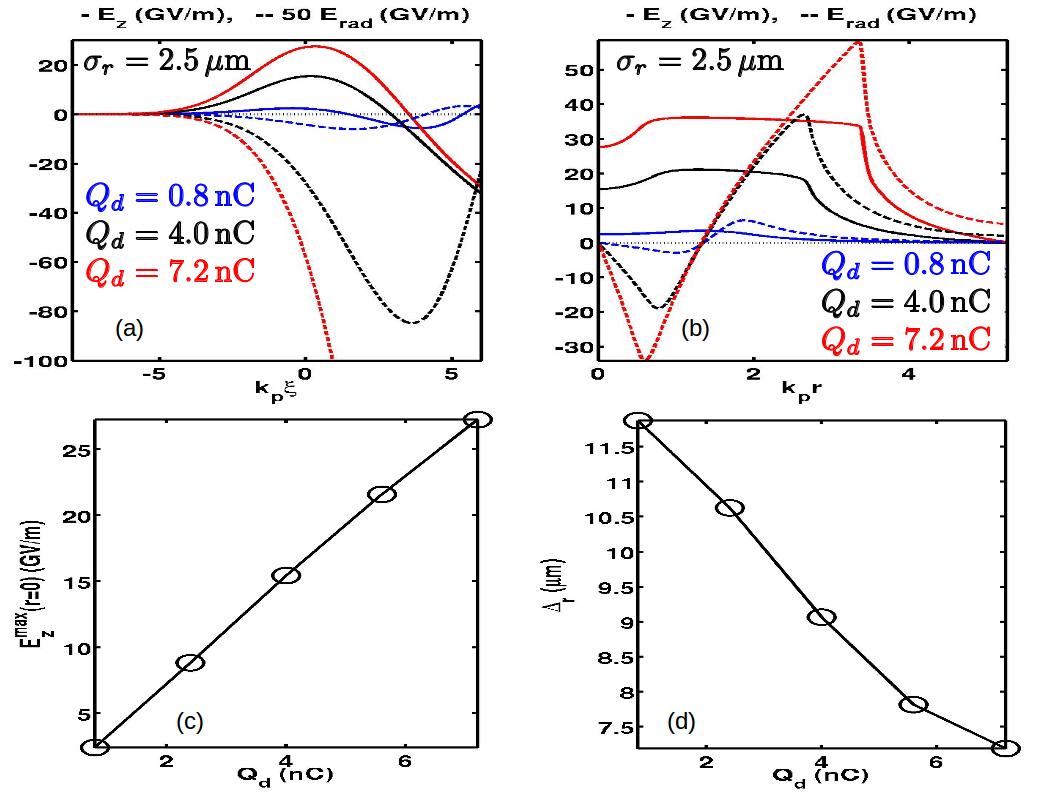}
 \caption{\label{fig:Ez_Erad_lineouts_Qd} Line-outs of Longitudinal
electric field $E_z$ and transverse field $E_{rad}=E_r-cB_{\theta}$ along the
axis
($r=0$; a) and radius (at axial locations where $E_z$ is maximum in the
favorable region for positron acceleration; b) for various values of
total charge in the hollow beam driver ($\sigma_r=2.5\mu$m, $r_0=7\sigma_r$).
Scaling of $E_z^{max}(r=0)$ (c) and $\Delta_r$ (d) with the total charge in
the hollow beam driver.} 
\end{figure}

The axial and radial line-outs of wake fields in Figs.
\ref{fig:Ez_Erad_lineouts_Qd}a and b show that  the accelerating
electric field increases  with the total charge $Q_d$ contained in the hollow
beam
driver. The maximum of the on-axis $E_z$ scales linearly with $Q_d$ (Fig.
\ref{fig:Ez_Erad_lineouts_Qd}c). The axial size of the favorable region for
positron acceleration 
increases only by a small length. This is unlike the favorable region for
positron acceleration in the back of the solid beam driver. In the latter case,
the
axial size of the favorable region shrinks with the energy content of the driven
plasma wave and thus increasing the charge in the beam does not improve the
efficiency. The focusing field driven by the hollow beam driver varies faster
with $\xi$ for larger values of $Q_d$. However, this variation of the focusing
field is still slower than that in the back of the solid beam driver. The
radial extent $\Delta_r$ (over which the focusing field is linearly varying
with radius) decreases with $Q_d$ but remains of the order of $k_p^{-1}$ for
up to $Q_d$=7.2 nC (Fig. \ref{fig:Ez_Erad_lineouts_Qd}d). The radial uniformity
of $E_z$ is degraded for high values
of charge in the beam.


Now, we show by simulations that a witness beam of positrons can be efficiently
accelerated in the wake fields generated by a hollow electron beam driver. For
this purpose we place a positron beam with a density profile given by, 
\begin{eqnarray}
 n_{witness}&=&n_w\exp\left[ 
 -\frac{r^2}{2\sigma_{r,w}^2}-\frac{(\xi-\xi_w)^2}{2\sigma_{z,w}^2}\right]
\label{eq:witness},
\end{eqnarray}
on the axis of an evolving hollow electron beam driver whose initial density
profile is given
by Eq. (\ref{eq:driver}). The background plasma has a uniform 
density, $n_p=5\times 10^{16}$ cm$^{-3}$ giving $k_p^{-1}=23.79 \,\mu$m, and is
modeled using 9 simulation particles per cell. The parameters for electron beam
driver are $\xi_0=0$, $\sigma_z=23.79 \,\mu$m $=k_p^{-1}$, $\sigma_r=4.76
\,\mu$m $=0.2 k_p^{-1}$, $r_0/\sigma_r=10$ and total charge $Q_d=-5.12$ nC
corresponding to $n_b=3.0$. The witness bunch parameters are
$\sigma_{z,w}=\sigma_{r,w}=4.76 \,\mu$m $=0.2\,k_p^{-1}$,  
$\xi_w=6.26 \,\mu$m and total charge in witness bunch $Q_w=13.58$ pC
corresponding to $n_w=1.0$. The electrons in the driver and positrons in
the witness beam have an initial energy
of 23 GeV and are modeled using 1.25 $\times 10^6$ and 6.25 $\times
10^5$ simulation particles, respectively.  The simulation domain size along
$\xi$ is approximately $262 \,\mu$m $\approx 11 \,k_p^{-1}$ with a grid
resolution $d\xi=0.52 \,\mu$m $=0.02 \,k_p^{-1}$ while along $r$ is 119 $\mu$m
$\approx 5 \,k_{p}^{-1}$ with a grid resolution of $dr\approx0.6\,\mu$m $=0.025
\,k_p^{-1}$. The driver and witness beams are propagated in the uniform plasma
in steps of  $ds\approx 4.75 \,\mu$m for a total
distance of 140 cm.

\begin{figure}
 \includegraphics[width=0.5\textwidth,height=0.4\textheight]
 {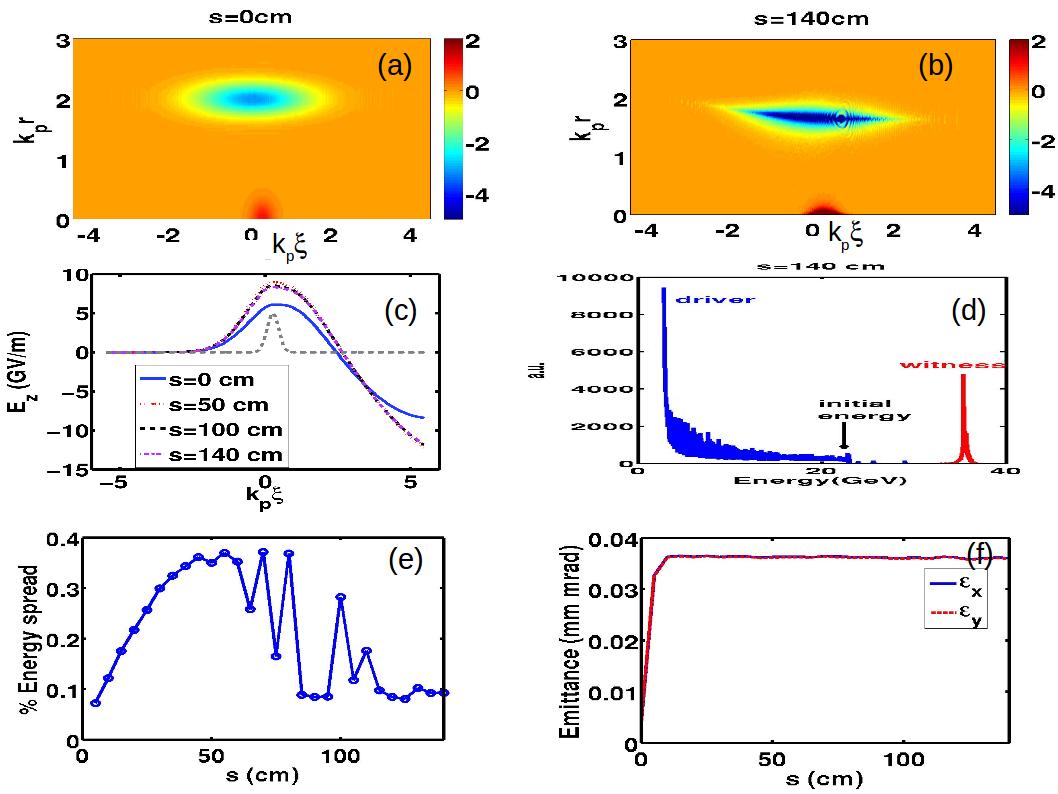}
 \caption{\label{fig:en_gain_spread} Charge densities of the driver (electrons)
and witness (positrons) beams before  the beams begin to propagate (a) and after
they have propagated a  distance $s=140$ cm (b). The color scale for charge
density is saturated at $-5.0$ and $2.0$. Line-outs of longitudinal electric
field $E_z$ (at various propagation distances) and initial profile of witness
(positrons) charge density (gray dashed line; in arbitrary units) along the
axis ($r=0$) (c). Energy spectrum of driver and witness beams after 140 cm of propagation in plasma (d). Energy spread in percentage (e) and normalized RMS $x$ and $y$ emittance (f) in witness beam as a function of propagation distance $s$.} 
\end{figure}

Figs. \ref{fig:en_gain_spread}a and \ref{fig:en_gain_spread}b shows  charge densities of the driver and witness beams before the beams begin to propagate (propagation distance, $s=0$) and after they have propagated a distance of 140 cm. 
During the propagation, the electron beam driver shifts radially inward and
reduces its radial spread, thereby reducing the effective values of 
$r_0$ and $\sigma_r$, respectively. This can be seen by comparing Figs.
\ref{fig:en_gain_spread}a and \ref{fig:en_gain_spread}b. As a result, the
longitudinal electric field $E_z$ also increases in and around the axial extent
of the witness beam. However, as shown in Fig. \ref{fig:en_gain_spread}c, $E_z$
saturates at slightly larger values by $s\approx50$ cm.  The energy gain of the
witness beam after propagating 140 cm is 12.4 GeV in Fig.
\ref{fig:en_gain_spread}d. The estimate of work done by the electric field can
be obtained from $W=e\int_0^{140 \mathrm{cm}}E_z^{peak}(s) ds\approx12.43$ GeV,
where $E_z^{peak}$ is the maximum value of $E_z$ on the axis. This estimate is 
very close to the energy gained by the witness beam. 
 
 The axial position of the witness beam was chosen to produce and sit
in a uniform region of $E_z$. This uniform region becomes slightly nonuniform
due to beam evolution. However it does not significantly affect 
the energy spread of the accelerated witness beam as can be seen in Fig.
\ref{fig:en_gain_spread}e. The percentage energy spread of the witness beam
defined as $(\Delta K_{fwhm}/K_{max})\times 100$  first increases and then
decreases with the propagation distance. Here $\Delta K_{fwhm}$ is the full
width at half maximum of the energy spectrum of the witness beam and  $K_{max}$
is the energy corresponding to the peak of the energy spectrum, Fig.
\ref{fig:en_gain_spread}d. The maximum value of the percentage energy spread
remains under 0.4. In Fig. \ref{fig:en_gain_spread}f, the
normalized RMS emittance of the witness beam, $\epsilon_x$ and $\epsilon_y$
(defined as $\epsilon_{x}=\sqrt{<x^2><p_x^2>-<xp_x>^2}/m_ec$ where $x$ and $p_x$
are the position and 
momentum of a witness particle, and similar definition for $\epsilon_y$),
does not increase much and remains close to its initial value during the beam's
propagation. 
 
Accelerating gradients for positrons in excess of 5 GV/m have been earlier
reported \cite{kimura2011,wang2008}.  
For appropriate beam parameters ($Q_d$, $r_0$ and $\sigma_r$), wake
fields generated in a plasma by a hollow electron beam driver can offer radially
uniform higher accelerating gradient and linear focusing field for positron
acceleration. The typical radial and axial sizes of the available positron
beams in experiments are approximately 10 $\mu$m and $< 100\, \mu$m,
respectively \cite{hogan2010}. For $n_0=2\times10^{17}$ cm$^{-3}$,  the
radial and axial sizes of the favorable region are approximately
$k_p^{-1}=12\,\mu$m
and $\lambda_p=2\pi/k_p=75\,\mu$m respectively, which are sufficiently large for
the placement of a positron witness beam of available sizes. The most attractive
feature of this scheme is the linear scaling of the longitudinal electric field
with the charge in the driver beam without compromising the size of the
favorable
region.
\begin{acknowledgments}
This work was supported by the US DoE grant number DESC0007970.
\end{acknowledgments}
%

\end{document}